\documentclass[conference]{IEEEtran}
\IEEEoverridecommandlockouts
\usepackage{cite}
\usepackage{enumitem}
\usepackage{amsmath,amssymb,amsfonts}
\usepackage{booktabs, multicol, multirow}
\usepackage{tablefootnote}
\usepackage[breakable]{tcolorbox}
\usepackage{bbding}
\usepackage{pifont}
\newcommand{\cmark}{\ding{51}}%
\newcommand{\xmark}{\ding{55}}%
\usepackage{algorithmic}
\usepackage{graphicx}
\usepackage{textcomp}
\usepackage{xcolor}

\def\BibTeX{{\rm B\kern-.05em{\sc i\kern-.025em b}\kern-.08em
    T\kern-.1667em\lower.7ex\hbox{E}\kern-.125emX}}
\begin{document}

\title{Learning Program Representations with a Tree-Structured Transformer
}

\author{\IEEEauthorblockN{Wenhan Wang, Kechi Zhang, Ge Li, Shangqing Liu, Anran Li, Zhi Jin, Yang Liu}
\and
\IEEEauthorblockN{Author}
\IEEEauthorblockA{\textit{dept. name of organization (of Aff.)} \\
 City, Country \\
 email address or ORCID}
}

\author{\IEEEauthorblockN{Wenhan Wang\IEEEauthorrefmark{2}, Kechi Zhang\IEEEauthorrefmark{1}, Ge Li\IEEEauthorrefmark{1}\textsuperscript{\textsection}, Shangqing Liu\IEEEauthorrefmark{2}, Anran Li\IEEEauthorrefmark{2}, Zhi Jin\IEEEauthorrefmark{1}\textsuperscript{\textsection}, Yang Liu\IEEEauthorrefmark{2}}
\IEEEauthorblockA{\IEEEauthorrefmark{1}Key laboratory of High Confidence Software Technologies (Peking University), Ministry of Education, \\
Institute of Software, EECS, Peking University, Beijing, China}
\IEEEauthorblockA{\IEEEauthorrefmark{2}Nanyang Technological University, Singapore}
\IEEEauthorblockA{Email: wenhan.wang@ntu.edu.sg, zhangkechi@pku.edu.cn, lige@pku.edu.cn, shangqin001@e.ntu.edu.sg, \\
anran.li@ntu.edu.sg, zhijin@pku.edu.cn, yangliu@ntu.edu.sg}
}

\maketitle

\begingroup\renewcommand\thefootnote{\textsection}
\footnotetext{Corresponding authors}
\endgroup

\begin{abstract}
Learning vector representations for programs is a critical step in applying deep learning techniques for program understanding tasks. Various neural network models are proposed to learn from tree-structured program representations, \emph{e.g.}, abstract syntax tree (AST) and concrete syntax tree (CST). However, most neural architectures either fail to capture long-range dependencies which are ubiquitous in programs, or cannot learn effective representations for syntax tree nodes, making them incapable of performing the node-level prediction tasks, \emph{e.g.}, bug localization. 
In this paper, we propose Tree-Transformer, a novel recursive tree-structured neural network to learn the vector representations for source codes. We propose a multi-head attention mechanism to model the dependency between siblings and parent-children node pairs. Moreover, we propose a bi-directional propagation strategy to allow node information passing in two directions, bottom-up and top-down along trees. In this way, Tree-Transformer can learn the information of the node features as well as the global contextual information.
The extensive experimental results show that our Tree-Transformer significantly outperforms the existing tree-based and graph-based program representation learning approaches in both the tree-level and node-level prediction tasks.

\end{abstract}


\section{Introduction}
\begin{figure*}[htbp]
\centering
\includegraphics[width=0.9\textwidth]{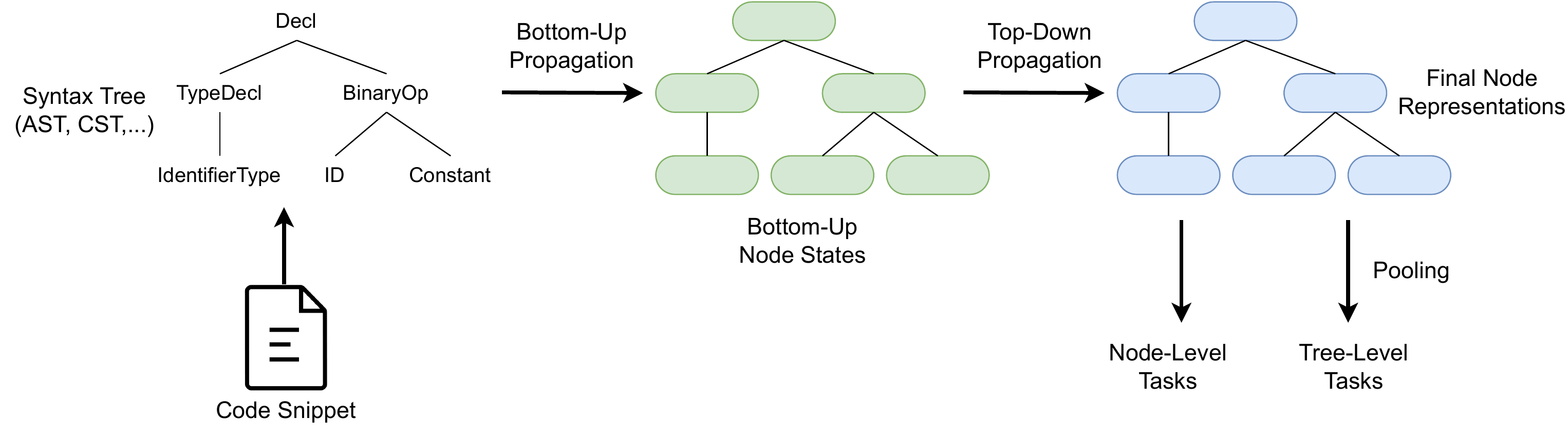} %
\caption{The overall pipeline of Tree-Transformer.}
\label{fig: propagation}
\vspace{-6mm}
\end{figure*}

The rapid development of software engineering applications is incurring enormous growth of code-related data, which makes ``Deep Learning (DL) for Software Engineering (SE)" particularly important. 
DL for SE have been improving software development in many application fields, such as clone detection \cite{wei2017supervised,wang2020detecting}, vulnerability detection \cite{zhou2019devign} and code summarization \cite{wan2018improving,zugner2021language}.
One major challenge for these DL for SE applications is how to represent source code to capture their syntactical and semantic information effectively. 

Existing DL for SE works have incorporated syntactic or semantic structure information, \emph{e.g.}, abstract syntax tree (AST) and data/control flow into the learning process through tree-structured neural networks~\cite{mou2016convolutional,zhang2019novel,wang2020modular,bui2021treecaps} or graph neural networks (GNN)~\cite{allamanis2018learning,liu2020retrieval,fernandes2019structured}. The tree-based approaches extract the syntax tree of the source code and use it as the input to the learning model, while graph-based methods use various program analysis techniques to convert the program into a graph and use it as the input to the model.
For example, Mou et al. \cite{mou2016convolutional} propose the tree-based convolutional neural network (TBCNN) and apply it to the program classification using ASTs. Allamanis et al. \cite{allamanis2018learning} convert programs to graphs by adding data flow edges to ASTs, and apply the GNNs to the program graphs to identify misused variables. 


Although the existing structured-based techniques have shown their advantages on various software engineering tasks, they still have the following limitations. For better illustration, we summarize the existing structured-based techniques in Table \ref{tbl:summary}. \textbf{First}, many graph-based approaches fail to capture the long-term dependencies in source code (see column ``Global" in Table \ref{tbl:summary}). Since the long-term dependencies are ubiquitous in programs, \emph{e.g.}, a statement that calls a variable may be far from the definition of that variable, capturing long-range dependencies is critical for learning code representations \cite{hellendoorn2019global,liu2020retrieval, liu2021graphsearchnet}. However, most GNNs only learn local dependencies within small neighbourhoods, due to their message-passing mechanism.
\textbf{Second}, many tree-structured neural models for code cannot learn useful node representations (see column ``Nodes" in Table \ref{tbl:summary}). When structured deep learning models are needed to solve certain tasks, \emph{e.g.}, identifying misused variables \cite{allamanis2018learning} or inferencing variable types \cite{allamanis2020typilus}, we need to make predictions based on (syntax tree or program graph) node representations. An effective node representation should contain the information about the node features as well as its contextual information. However, since the node information in many tree-structured neural networks is propagated in a uni-directional bottom-up manner, these models cannot obtain sufficient context information for low-level tree nodes, especially for leaf nodes \cite{tai2015improved,mou2016convolutional,bui2021treecaps}. 
\textbf{Third}, most existing structured-based approaches ignore the order information in programs (see column ``Order" in Table \ref{tbl:summary}). Since the programs are highly ordered, \emph{e.g.}, program statements are executed in specific orders according to their control flow, whereas the message passing mechanism of GNN is permutation invariant and cannot capture the order information. 
\textbf{Last}, the existing graph-based methods and some tree-based approaches require additional processing of the syntax tree, which brings large overhead and loses the original structural information of the syntax tree (see column ``Original" in Table \ref{tbl:summary}). 
For example, building program graphs often requires experts to utilize various static analysis techniques \cite{allamanis2018learning,zhou2019devign}, and some of which are not adaptable to different program languages or program understanding tasks. 


To tackle the issues above, in this paper, we propose Tree-Transformer, a transformer-based approach to learning program representations for various downstream tasks. To model the global dependencies for all syntax tree nodes, we propose a multi-head attention-based bidirectional propagation method in opposite directions. 
The bidirectional propagation consists of two Tree-Transformer units, a bottom-up unit, and a top-down unit. The bottom-up unit aims to aggregate the contextual information from leaves to the root node 
, while the top-down unit aims to distribute the root information to its descendants. 
In this way, each node can capture the contextual information from all other nodes, which enables Tree-Transformer to learn long-range dependencies and meaningful node representations. Moreover, we adopt the position encoding mechanism of Transformer to realize the learning of sibling order information. 
Our contributions are summarized as follows: 

\begin{itemize}[leftmargin=*]
    \item We propose Tree-Transformer, a recursive tree-structured neural network which formulates the parent-child and sibling relations on the trees with multi-head attention to learn vector representations for program syntax trees.
    \item We design a bottom-up and top-down bidirectional information propagation method on the Tree-Transformer to capture the global dependencies between any pair of syntax tree nodes and enable node-level prediction where previous tree-structured neural networks fail.
    \item We have conducted extensive experiments on tree-level and node-level prediction tasks on program syntax trees. The experimental results demonstrate  that our approach significantly outperforms existing tree-structured neural networks, \emph{e.g.}, Tree-LSTM, and graph neural networks, \emph{e.g.}, GIN. For the node-level prediction task, we further design a wrong operator localization and repair task to evaluate the advantages of Tree-Transformer. 
    
\end{itemize}

\begin{table}[!t]
  \centering
  \caption{A comparison of existing tree/graph-based program representation learning techniques.}
    \begin{tabular}{lcccc}
    \toprule
     Approach & Global & Nodes & Order & Original \\
    \midrule
    GNN+AST & \xmark & \cmark & \xmark & \cmark\\
     Child-Sum Tree-LSTM~\cite{tai2015improved} & \cmark & \xmark & \xmark & \cmark\\
     N-ary Tree-LSTM \cite{tai2015improved} & \cmark & \xmark & \cmark & \xmark\\
     TBCNN \cite{mou2016convolutional} & \xmark & \xmark & \cmark & \cmark\\
     TreeCaps \cite{bui2021treecaps} & \cmark & \xmark & \cmark & \cmark\\
     Code2Vec \cite{alon2019code2vec} & \xmark & \cmark & \xmark & \xmark\\
     Code2Seq \cite{alon2018code2seq} & \xmark & \cmark & \xmark & \xmark\\
     ASTNN \cite{zhang2019novel} & \cmark & \xmark & \cmark & \xmark\\
     Tree-PE \cite{shiv2019novel} & \cmark & \cmark & \cmark & \xmark\\
     Code Transformer \cite{zugner2021language} & \cmark & \cmark & \cmark & \xmark\\
     TreeBERT \cite{jiang2021treebert} & \cmark & \cmark & \cmark & \xmark\\
     \textbf{Tree-Transformer(Ours)} & \cmark & \cmark & \cmark & \cmark\\
    \midrule
     GNN+Augmented AST \cite{allamanis2018learning} & \xmark & \cmark & \cmark & \xmark\\
     Devign \cite{zhou2019devign} & \xmark & \cmark & \cmark & \xmark\\
     GREAT \cite{hellendoorn2019global} & \cmark & \cmark & \cmark & \xmark\\
     HPG+HGT \cite{zhang2022learning} & \xmark & \cmark & \cmark & \xmark\\
    \bottomrule
    \end{tabular}%
  \label{tbl:summary}%
\end{table}%

\section{Motivating Example}
Here, we take a specific example to demonstrate the weakness of existing program representation learning approaches on syntax trees, and the design motivation of Tree-Transformer.

\begin{figure*}[htbp]
\centering
\includegraphics[width=0.9\textwidth]{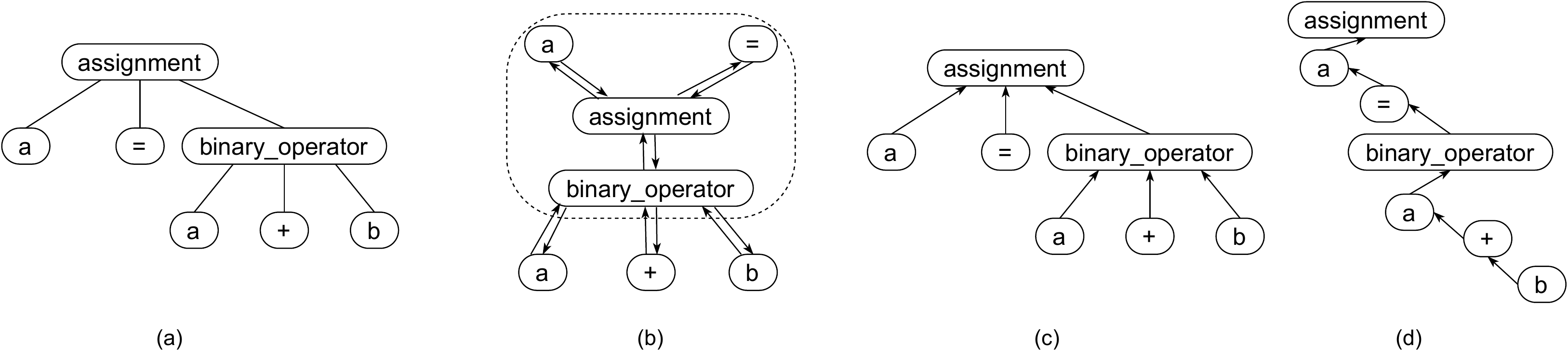} %
\caption{An example of existing tree-structured and graph neural networks for learning on abstract syntax trees.}
\label{fig: motivation}
\vspace{-4mm}
\end{figure*}

Figure \ref{fig: motivation} (a) demonstrate a typical abstract syntax tree. When using graph neural networks on ASTs, node information is propagated bi-directionally along AST edges (see Figure \ref{fig: motivation} (b)). Most GNN models adopt the message passing mechanism \cite{gilmer2017neural}, \emph{i.e.}, each node only receives information from its $k$-th local neighborhoods ($k$ is the number of GNN layers). For example, the dashed box in Figure \ref{fig: motivation} (b) is the 1-hop local neighborhood of node ``assignment". Consequently, such GNN methods lack the capability of capturing longer dependencies. 

In most tree-structured neural networks\cite{tai2015improved,mou2016convolutional,ahmed2019you}, node information is propagated in a uni-directional bottom-up fashion (see Figure \ref{fig: motivation} (c)). This means that each node receives context information from its descendants. Low-level nodes, especially leaf nodes, cannot receive context from an adequate number of nodes, so the representations learned for these nodes are unlikely to perform well on node-level prediction tasks. For example, in Figure \ref{fig: motivation} (c), the leaf node ``b" cannot receive any other information other than itself. 

Another issue for GNNs and some tree-structured neural networks is that they cannot naturally handle the node order information in ASTs. To alleviate this issue, some works adopt order-sensitive neural networks \cite{wei2017supervised,wan2018improving,wan2019multi,shiv2019novel}, which are mainly built for N-ary trees, especially binary trees. However, an AST node may have an arbitrary number of children, and converting ASTs to N-ary trees will destroy the original tree structure information. Figure \ref{fig: motivation} (d) shows an AST converted into a binary tree with left-child right-sibling (adopted by \cite{shiv2019novel}), and we can obviously see that the original tree structure is vastly changed.


\section{Approach}
\begin{figure*}[t]
\centering
\includegraphics[width=0.95\textwidth]{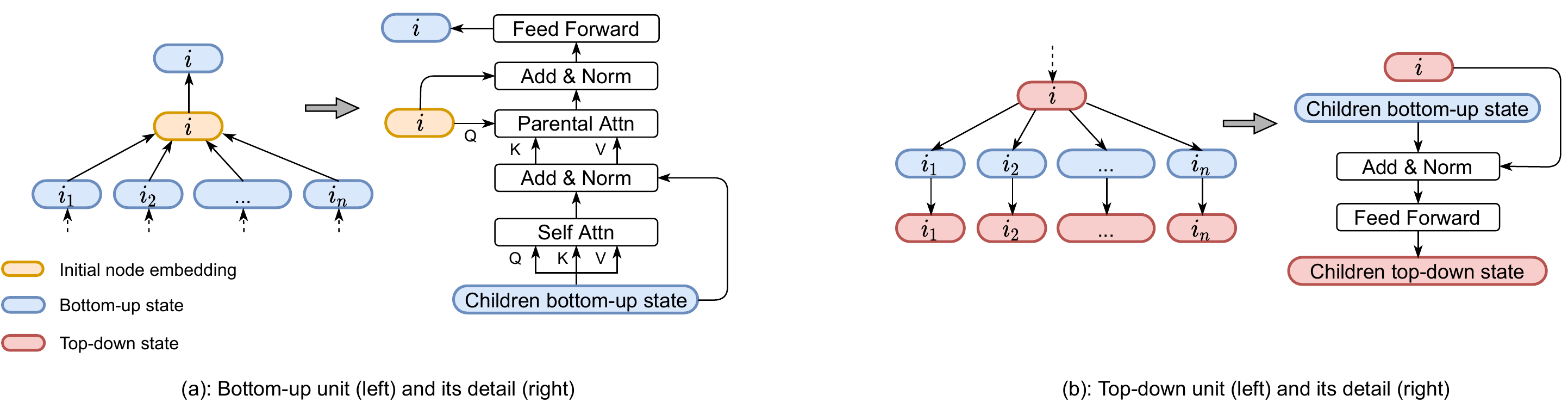} 
\caption{The detailed architecture of Tree-Transformer units. (a): Bottom-up Tree-Transformer unit. (b): Top-down Tree-Transformer unit.}
\label{fig:propagation-unit}
\vspace{-6mm}
\end{figure*}

As shown in Figure 1, the Tree-Transformer consists of two consecutive steps: 

\begin{itemize}[leftmargin=*]
    \item Bottom-up propagation: a bottom-up Tree-Transformer unit propagates the node messages recursively from children to parents to obtain the bottom-up states of nodes.
    \item Top-down propagation: a top-down Tree-Transformer unit distributes the learned contextual information from the parent node to their children.
\end{itemize}

After the bi-directional propagation of Tree-Transformer, the obtained top-down node states can be treated as final node representations, and further utilized for node-level prediction tasks (e.g. bug localization). We can also pool the node representations learned by Tree-Transformer into a single vector for tree-level prediction tasks (e.g. program classification). In the following we will describe the bi-directional propagation of Tree-Transformer in details.

\subsection{Bottom-up Propagation}
Formally, a code snippet $C$ can be parsed into an AST $\mathcal{T}=(\mathcal{V}_{leaf}, \mathcal{V}_{non})$, where $\mathcal{V}_{leaf}$ and $\mathcal{V}_{non}$ denote the set of AST leaf nodes and non-leaf nodes respectively. Any node $i \in  \mathcal{V}_{non}$ connects with a set of child nodes $\mathrm{c}_{i} = \{i_1, i_2, ..., i_n \}$ where $n$ is the number of child nodes for $i$. In ASTs, different nodes may have different numbers of $n$. 
For any node $v \in \mathcal{T}$, we utilize a learnable embedding matrix $\mathbf{E}$ to obtain the initial node embedding $\boldsymbol e_v \in \mathbb{R}^{d}$, where $d$ is the dimension of node embeddings. 


To gather the children information for the non-leaf node $i$, we design the bottom-up propagation based on the multi-head attention~\cite{vaswani2017attention}. The architecture of the bottom-up Tree-Transformer unit is shown in Figure~\ref{fig:propagation-unit} (a). 

In multi-head attention, each attention head can be formulated as:
\begin{equation}
    Attention(Q,K,V) = softmax(\frac{QK^{T}}{\sqrt{d}})V , \label{attn}
\end{equation}

Where $Q$, $K$, $V$ denote the query, key and value, and $d$ is the dimension of key vectors.

The bottom-up Tree-Transformer unit obtain the bottom-up states of nodes in a bottom-up manner similar to recursive neural networks \cite{goller1996learning}, i.e., a node $i$'s bottom-up state $\boldsymbol h_{i\uparrow}$ is updated from its initial node embedding $\boldsymbol e_i$ and its children's bottom-up states $\boldsymbol H_{\mathrm{c}_{i}\uparrow}=(\boldsymbol h_{i_1\uparrow}, \boldsymbol h_{i_2\uparrow}, ..., \boldsymbol h_{i_n\uparrow})$. If the child nodes of $i$ are leaf nodes, i.e., $\{i_1, ..., i_n \} \subseteq  \mathcal{V}_{leaf}$, $\boldsymbol H_{\mathrm{c}_{i}\uparrow}$ are equal to their leaf node embeddings i.e., $\boldsymbol H_{\mathrm{c}_{i}\uparrow} = (\boldsymbol e_{i_1}, \boldsymbol e_{i_2}, ..., \boldsymbol e_{i_n})$. 

In a bottom-up Tree-Transformer unit, we first apply a fraternal self-attention $\mathrm{MultiHead}_{f\uparrow}$ on $\boldsymbol H_{\mathrm{c}_{i}\uparrow}$ to model the sibling dependency between nodes in $\mathrm{c}_{i}$. In the fraternal attention, the query, key and value all come from the children node bottom-up state sequence. Then we use a parental multi-head attention $\mathrm{MultiHead}_{p}$ to capture the dependency between $i$ and its children. In the parental attention of $i$, the query is its initial node embedding $\boldsymbol e_i$, and the key/value are the output of the fraternal attention on $i$'s children. The output of the parental attention is thereafter used to update the bottom-up state of $i$. Like the sequential transformer model, each multi-head attention operation in the Tree-Transformer unit is followed by a layer normalization and residual connection.
Finally, we use a position-wise feed-forward layer same as \cite{vaswani2017attention} calculates the bottom-up state. Formally, the bottom-up Tree-Transformer unit calculates ${\boldsymbol h}_{i \uparrow}$, the bottom-up state of $i$ by:

\begin{align}
    \boldsymbol H'_{\mathrm{c}_{i}\uparrow}&=\mathrm{MultiHead}_{f\uparrow}(\boldsymbol H_{\mathrm{c}_{i}\uparrow}, \boldsymbol H_{\mathrm{c}_{i}\uparrow}, \boldsymbol H_{\mathrm{c}_{i}\uparrow}) \\
    \boldsymbol H'_{\mathrm{c}_{i}\uparrow}&=\mathrm{LayerNorm}(\boldsymbol H'_{\mathrm{c}_{i}\uparrow}+\boldsymbol H_{\mathrm{c}_{i}\uparrow}) \\
    \mathrm{\boldsymbol A_{\uparrow}}&=\mathrm{MultiHead}_{p}(\boldsymbol e_i, \boldsymbol H'_{\mathrm{c}_{i}\uparrow}, \boldsymbol H'_{\mathrm{c}_{i}\uparrow})  \\
    \boldsymbol A'_{\uparrow}&=\mathrm{LayerNorm}(\mathrm{\boldsymbol A}+\boldsymbol e_i)  \\
    {\boldsymbol h}_{i \uparrow}&=\mathrm{LayerNorm}(\mathrm{FFN}_{\uparrow}(\boldsymbol  A'_{\uparrow})+ \boldsymbol A'_{\uparrow}) 
\end{align}

To capture the sibling order information in syntax trees, we apply position encodings in the fraternal self-attention $\mathrm{MultiHead}_{f\uparrow}$. This allows our model to handle sibling orders for arbitrary trees, while many existing order-sensitive models on trees \cite{tai2015improved,shiv2019novel} only work on trees with a fixed branching factor (N-ary trees).

Different from the standard position encoding of Transformer~\cite{vaswani2017attention}, we adopt the Untied Positional Encoding (TUPE)~\cite{ke2021rethinking} in the bottom up fraternal attention $\mathrm{MultiHead}_{f\uparrow}$. In traditional position encoding, the token embeddings and position embedding vectors are added before the multi-head attention, this makes the model cannot distinguish token information and position information after these two are mixed up. The TUPE position encoding unties the position information from the token information, this enables the Transformer model to learn more precise position information without the interference of input tokens. When we use TUPE in Tree-Transformer, the output $\boldsymbol H'_{\mathrm{c}_{i}\uparrow}=(z_{1},z_{2},...,z_{n})$ in Equation (2) is computed by:

\begin{align} 
    \boldsymbol z_{m}&=\sum_{j=1}^{n}\frac{\mathrm{exp}(\boldsymbol \alpha _{mj})}{\sum_{j'=1}^{n}\mathrm{exp}(\boldsymbol \alpha _{mj'})}(\boldsymbol h_{i_j\uparrow} \boldsymbol W^{V}) \\
    \alpha _{mj}&=\frac{1}{\sqrt{2d}}(\boldsymbol h_{i_m\uparrow}\boldsymbol W^{Q})(\boldsymbol h_{i_j\uparrow}\boldsymbol W^{K})^{T}+\frac{1}{\sqrt{2d}}(\boldsymbol p_{m} \boldsymbol U^{Q})(\boldsymbol p_{j}\boldsymbol U^{K})^{T}
\end{align}
Where $\boldsymbol W^{Q},\boldsymbol W^{K},\boldsymbol W^{V} \in \mathbb{R}^{d \times d}$ are query/key/value projection matrices and $\boldsymbol U^{Q},\boldsymbol U^{K} \in \mathbb{R}^{d \times d}$ are (absolute) position projection matrices for queries and keys. $\boldsymbol p \in \mathbb{R}^{d}$ are the fixed positional embedding vectors.





\subsection{Top-down Propagation}
After bottom-up propagation, Tree-transformer obtains the bottom-up states of all nodes in AST i.e., $\{\boldsymbol h_{v \uparrow} | \forall v \in \mathcal{V}\}$. Tree-Transformer then performs the top-down propagation, which is shown in Figure~\ref{fig:propagation-unit} (b). When the top-down propagation is finished, each node can obtain the contextual information from all other nodes, thus enables to capture the global dependency which is missed in most existing tree/graph-structured neural networks. 

The top-down Tree-Transformer unit uses the state $\boldsymbol h_{i\downarrow}$ of a single node $i$ to simultaneously update its children's bottom-up states $\boldsymbol H_{\mathrm{c}_{i}\uparrow}=\{\boldsymbol h_{i_1\uparrow}, \boldsymbol h_{i_2\uparrow}, ..., \boldsymbol h_{i_n\uparrow}\}$ into top-down states $\boldsymbol H_{\mathrm{c}_{i}\downarrow}= \{\boldsymbol h_{i_1\downarrow}, \boldsymbol h_{i_2\downarrow}, ..., \boldsymbol h_{i_n\downarrow}\}$. If $i$ is the root node, $\boldsymbol h_{\mathrm{root} \downarrow}=\boldsymbol h_{\mathrm{root} \uparrow}$.
In the top-down parental attention, we aim to use a top-down parental attention to pass information from parent to children. In contrast to the bottom-up parental attention, in the top-down attention, children states $\boldsymbol H_{\mathrm{c}_{i}\uparrow}$ are used as query, and their parent top-down state $\boldsymbol h_{i\downarrow}$ as key/value.
This is not a common case for multi-head attention, because when the length of key/value is 1, the softmax function over keys/values is meaningless (it will always output 1). So we make a slight change and simplification to the top-down parental ``attention" function. Instead of computing attention scores, we directly add the top-down states of the parent node to all its children (this acts similar to a residual connection, the attention is omitted). 
The calculation process of the top-down unit is demonstrated below:

\begin{align}
    \boldsymbol A_{\downarrow}&=\mathrm{LayerNorm}(\boldsymbol 1 \cdot {\boldsymbol h}_{i \downarrow} + \boldsymbol H_{\mathrm{c}\uparrow})  \\
    \boldsymbol H_{\mathrm{c}\downarrow} &=\mathrm{LayerNorm}(\mathrm{FFN}_{\downarrow}(\boldsymbol H'_{\mathrm{c}\downarrow})+ \boldsymbol H'_{\mathrm{c}\downarrow}) 
\end{align}

After top-down propagation, Tree-Transformer obtains the top-down node states $\{\boldsymbol h_{v \downarrow} | \forall v \in \mathcal{V}\}$ which are used as the final node representations. 

By combining bottom-up and top-down propagation, Tree-Transformer is capable for modeling dependencies between any node pairs along paths with arbitrary lengths. On the contrary, although traditional Transformers can capture global dependencies, they can only model paths with a maximum length (the number of Transformer layers).


\subsection{Calculating Tree-Level Representation for Program-Level Prediction}
Different from previous (recursive) tree-structured neural networks, which use the state of root nodes as representation vector for trees, we use a pooling function over the top-down states for all nodes $\{v|v \in \mathcal{T}\}$ to obtain the final representation of a tree $\mathcal{T}$. We adopt the global attention pooling function proposed in \cite{li2016gated}:

\begin{equation}
    {\bf h}_{\mathcal{T}}=\sum_{v\in \mathcal{T}} \mathrm{softmax}(\boldsymbol W_{\mathrm{gate}}{\boldsymbol h}_{v\downarrow})\odot {\boldsymbol h}_{v\downarrow}
\end{equation}
$\boldsymbol W_{\mathrm{gate}}$ is a weight of $\mathbb{R}^{d}$, $\odot$ is the element-wise multiplication and ${\bf h}_{\mathcal{T}}$ can be utilized as the tree-level prediction.

\section{Experimental Setup}

In this section, we first introduce the selected tasks and baselines for evaluation, then present the settings for conducting out experiments.
\subsection{Tasks and Datasets}
We select three different tasks to evaluate the learning capacity of Tree-Transformer on learning tree-level and node-level representations. The detailed statistics of all datasets are listed in Table~\ref{tbl:dataset}.

\begin{table*}[htbp]
  \centering
  \caption{Basic statictics of the datasets we use in this paper. For CodeNet datasets, their vocabulary size is the size of their token vocabulary plus type vocabulary.}
    \begin{tabular}{lccccccc}
    \toprule
    & POJ & Java250 & Python800 & C++1000 & C++1400 & Wrong Operator & Type Inference\\
    \midrule
    Train samples & 36,400 & 45,000 & 144,000 & 300,000 & 252,000 & 155,628 & 608,156\\
    Validation samples & 5,200 & 15,000 & 48,000 & 100,000 & 84,000 & 16,868 & 24,424\\
    Test samples & 10,400 & 15,000 & 48,000 & 100,000 & 84,000 & 86,231 & 27,870\\
    \midrule
    Avg. nodes & 189.58 & 339.33 & 232.27 & 376.90 & 472.04 & 222.39 & 652.53\\
    Avg. children per node & 1.90 & 3.05 & 2.77 & 3.09 & 3.12 & 2.84 & 3.03\\
    Avg. depth & 13.32 & 17.26 & 14.48 & 15.46 & 16.43 & 13.28 & 17.09\\
    Vocabulary & 44 & 222 & 161 & 346 & 346 & 286,456 & 100,128 (manually set)\\
    \bottomrule
    \end{tabular}%
  \label{tbl:dataset}%
\end{table*}%

\subsubsection{\textbf{Program Classification}} In this task, we use Tree-Transformer to classify program ASTs based on the functionalities they implemented. We select this task to measure the ability of Tree-Transformer on learning tree-level representations. Specifically, we use two different datasets for evaluation. The first dataset is POJ algorithm classification dataset \cite{mou2016convolutional}, which has been widely adopted to evaluate the capability of program representation models. POJ dataset contains 104 classes of C programs from student programming platforms. In our experiment, we follow the AST parsing process of \cite{mou2016convolutional}: using pycparser\footnote{https://github.com/eliben/pycparser} to parse the functions to obtain ASTs. The second one is the CodeNet dataset \cite{puri2021project}, which contains over 14M code samples from two open judge platforms. Here we use the code classification benchmarks, which include four classification datasets in three programming languages: Java250, Python800, C++1000, and C++1400. These four datasets contain programs in 250, 800, 1000, and 1400 classes, respectively. Since CodeNet has already provided simplified parse trees (SPT) for those benchmarks, we directly use them for evaluation. CodeNet SPTs are generated by the ANTLR4 \cite{parr2013definitive} parser with a series of post-processing steps, including removing internal nodes with only one child. Notice that in both program classification datasets, the identifier names are discarded in syntax trees because we aim to perform classification on the algorithms alone without the additional information of identifier names. In open judge platforms, programmers tend to name identifiers according to the description of the programming problems, and the naming patterns may bring additional guidance for the program classifiers.

\subsubsection{\textbf{Wrong Operator Localization and Repair}} In order to evaluate Tree-Transformer on node-level prediction, we propose a novel tree-based wrong operator localization/repair task. Given the syntax tree of a code snippet with an erroneous binary operator (e.g., changing ``+" into ``-"), this task requires a model to locate the position of the misused operator node among all binary operator nodes, and predict the correct operator for this position. We synthesize a new dataset from the wrong operator detection dataset, released by CuBERT~\cite{kanade2020learning}. The original dataset 
is built for the binary classification between correct and buggy code snippets. To enable the localization and repair task, we only keep code snippets with more than one binary operator. On average, each code snippet in our dataset contains 5.98 binary operators.
For this dataset, we use tree-sitter\footnote{https://tree-sitter.github.io/tree-sitter/} to parse source code into concrete syntax trees (CST). CSTs contains more nodes than ASTs, mainly including brackets and punctuation.

\subsubsection{\textbf{Type Inference}} In this task, we utilize Tree-Transformer to classify the type of identifiers in TypeScript, a dynamically-typed language. Similar to wrong operator localization/repair, this task can also be seen as a prediction task on syntax tree nodes. For this task, we employ the public type inference dataset ManyTypes4TypeScript \cite{jesse2022many}, which consists of 13,953 public Github projects. We use the tree-sitter TypeScript parser to parse code snippets into CSTs and adopt a linear classifier on identifier nodes for predicting identifier types. We follow the original settings of ManyTypes4TypeScript and choose the vocabulary of identifier types as 50,000, so this task is a node classification task with 50,000 classes.

\subsection{Compared Baselines}
We compare our approach against existing tree-structured and graph-structured neural networks. We further compare with some transformer-based models built for programming languages. To sum up, we choose the following models as baselines:
\begin{itemize}[leftmargin=*]
    \item \textbf{Tree-structured neural networks}. We compare Tree-Transformer with Tree-LSTM \cite{tai2015improved}, TBCNN \cite{mou2016convolutional} and TreeCaps\cite{bui2021treecaps}. 
    As program syntax trees can have arbitrary branching factors, we use Child-Sum Tree-LSTM as our baseline. 
    For TreeCaps, we use the variable-to-static (VTS) version in our experiments.  
    
    \item \textbf{Graph neural networks}. We choose graph convolutional network (GCN) \cite{kipf2017semi}, graph isomorphism network (GIN) \cite{xu2019how}, and gated graph neural network (GGNN) \cite{li2016gated} as our baselines. 
    We evaluate the GNN baselines with two different sets of inputs: the original syntax tree and syntax tree augmented with NextToken edges (connect a terminal token node to the next terminal)\cite{allamanis2018learning,puri2021project}. 
    
    \item \textbf{Transformer-based models}. We choose two tree/graph-based Transformer models as our baselines: Tree-PE \cite{shiv2019novel} and GREAT \cite{hellendoorn2019global}. As Tree-PE can only handle N-ary trees, we convert the input trees to 10-ary trees for program classification and 15-ary for wrong operator localization and repair. As GREAT usually takes graphs with multiple edge types as inputs, we use syntax trees with NextToken edges as its inputs. 
    We also compare our approach with sequential Transformer models on source code token sequences. Our sequential baselines including a vanilla Transformer \cite{vaswani2017attention} and pre-trained models: CodeBERT~\cite{feng2020codebert} and C-BERT~\cite{buratti2020exploring}. Different from tree/graph-based approaches, in sequence-based approaches, we follow previous works \cite{puri2021project} and use the token sequences of the original code, which means the identifier names are kept.
\end{itemize}
 

\subsection{Experimental Settings}
\subsubsection{General Settings}
We set the Tree-Transformer node embedding dimension to 128 for POJ and 256 for other datasets. The number of attention heads is set to 4. 
We train our models with an Adam optimizer with a default learning rate of 0.002 and a warm-up phase of 2,000 steps. We implement Tree-Transformer with DGL \cite{wang2019deep} to enable efficient batching ,and ran our experiments on a NVIDIA RTX 8000 GPU with 48GB memory.

\subsubsection{Settings for Program Classification}
For the POJ dataset, we follow previous works \cite{bui2021treecaps} and split the dataset into train/validation/test sets by the ratio of 7:1:2. For CodeNet, the train/validation/test ratio is 3:1:1. Since each node in CodeNet SPTs contains two parts of information: parsing rules and tokens, thus we concatenate the token embeddings and the parsing rule embeddings as the initial node embeddings.
For GNN baselines, we adopt the same pooling function as Tree-Transformer. For Tree-LSTM, we employ two different approaches to acquire the representation vectors for trees: using the root node's hidden state or using the same attention pooling as our model. For TreeCaps, we follow its original setting and use its ``code capsule" to compute the probabilities of output classes.

\subsubsection{Settings for Wrong Operator Localization and Repair}
Locating the wrong operator node is achieved by learning a pointer pointing to a single node in a tree, which is the same as previous works on localization and repair tasks~\cite{vasic2019neural,hellendoorn2019global}. Unlike \cite{vasic2019neural}, which also uses a pointer for the repair task, We treat this step as a node classification task: a classifier predicts the label of the repair operator within the set of all binary operators. In our dataset, the operator set $OPs=\{-, +, *, \%, >, ==, \textrm{or}, <, /, \textrm{and}, >=, <=, !=, \textrm{in}, \textrm{is}, \textrm{is not}, \textrm{not in}\}$. We sum up the localization loss and the repair loss as the training loss for this task.

In this task, all the wrong operator nodes are located on the leaf nodes in CSTs. However, the tree-structured neural networks, e.g., Tree-LSTM, follow a bottom-up manner to propagate information, which means that the leaf nodes cannot receive the information from other nodes and cannot learn well-contextualized node representations. Thus we directly omit these tree-structured baselines and only utilize graph-based models for comparison.


\subsubsection{Settings for Type Inference}
The original ManyTypes4TyprScript dataset is created for token-base type inference on code token sequences, which cannot be directly converted to tree-base data without any data loss. In the parsing step, we remove code snippets that cannot be correctly parsed by tree-sitter and CSTs with larger than 5000 nodes, thus resulting in a slightly smaller dataset than the original one. The vocabulary size for CST terminal token nodes is manually set to 100,000 for this task. Similar to wrong operator localization and repair, we only compare with baselines that are capable of node-level predictions.

\section{Experimental Results}
In this section, we present and analyze our experiment results to address the following research questions:

\begin{itemize}
    \item \textbf{RQ1}: How does Tree-Transformer perform on syntax tree-level prediction tasks?
    
    \item \textbf{RQ2}: How does Tree-Transformer perform on node-level prediction tasks?
    
    \item \textbf{RQ3}: How does each component of Tree-Transformer contribute to our performances?
\end{itemize}

\subsection{RQ1: How does Tree-Transformer perform on syntax tree-level prediction tasks?}
Table~\ref{tbl:baselines} shows the classification results on CodeNet and POJ datasets. We can see that on all five datasets, Tree-Transformer outperforms the tree-structured and graph-structured baselines by a significant margin. This highlights the effectiveness of modeling global dependencies along trees with multi-head attention. When using the attention pooling the same as Tree-Transformer, Tree-LSTM does not show significant improvements, suggesting that our improvements over Tree-LSTM mainly come from our model design rather than the pooling strategy.
Moreover, Tree-Transformer outperforms all of our graph-structured baselines (including GNNs and GREAT), even when they are integrated with additional NextToken edges. Although adding additional edges to syntax trees (see rows ``GNN+Graph") can marginally improve the performances of program classification, it still cannot overcome the inherent weaknesses of graph-based models. We also include a heterogeneous graph-based model (HPG+HGT) \cite{zhang2022learning} for comparison. Although HPG+HGT outperforms (homogeneous) GNN baselines by introducing additional node and edge type information, it still cannot compete against our Tree-Transformer.

From the results, we notice that Tree-LSTM outperforms the GNN baselines when the given inputs are trees. Although the research interest in tree-structured neural networks is undermined by the advance of GNNs, tree-structured models are still competitive and should not be ignored.
When compared with large-scale pre-trained Transformers, we find that Tree-Transformer is still competitive. The average accuracies of C-BERT and CodeBERT on CodeNet is lower than ours, although for some datasets, for example, Java250 and Python800, they have a higher performance than ours. Compared with pre-trained baselines, which require massive data for pre-training, Tree-Transformer is much more light-weight, and this further confirms the effectiveness of our model. 

\begin{table*}[htbp]
   \centering
   \caption{Program classification accuracy(\%) on CodeNet and POJ datasets.}
     \begin{tabular}{lccccc|c}
     \toprule
     & \multicolumn{1}{l}{Java250} & Python800 & C++1000 & C++1400 & Overall & POJ\\
     \midrule
     GCN & 89.06 & 91.81 & 93.54 & 92.89 & 91.83 & 93.93\\
     GIN & 90.76 & 93.17 & 95.54 & 94.50 & 93.49 & 95.76\\
     GGNN & 88.46 & 89.92 & 89.75 & 88.01 & 89.04& 93.22\\
     \midrule
     Tree-LSTM (root) & 93.19 & 93.95 & 95.79 & 95.20 & 94.53& 94.70\\
     Tree-LSTM (attention) & 93.71 & 93.83 & 95.79 & 95.24 & 94.64& 94.95\\
     TBCNN & 90.32 & 91.10 & 93.17 & 93.03 & 91.91 & 94.15\\
     TreeCaps & 91.42 & 90.26 & 93.55 & 93.24 & 92.12& 95.88\\
     Tree-PE~\cite{shiv2019novel} & 91.65 & 91.11 & 92.30 & 88.97 & 91.01 & 94.19 \\
     \midrule
     GREAT & 93.15 & 93.30 & 93.46 & 93.72 & 93.41 & 92.64\\
     GCN+Graph \cite{puri2021project} & 92.70 & 93.82 & 95.76 & 95.26 & 94.39 & 95.57 \\
     GIN+Graph \cite{puri2021project} & 93.26 & 94.17 & 96.34 & 95.95 & 94.93 & 95.96\\
     GGNN+Graph & 93.61 & 92.23 & 91.72 & 92.48 & 92.51 & 94.80 \\
     HPG+HGT \cite{zhang2022learning} \tablefootnote{HPG \cite{zhang2022learning} can only build program graphs for Java and Python.} & 93.95 & 94.99 & - & - & - & -\\
     \midrule
     \textbf{Tree-Transformer (Ours)}   & 95.32 & 95.30 & \textbf{97.11} & \textbf{96.98} & \textbf{96.18} & \textbf{96.12}\\
     \midrule
     Transformer & 93.49 & 93.99 & 89.93 & 67.87 & 86.32 & 88.13\\
     C-BERT \cite{puri2021project} & \textbf{97.40} & \textbf{97.09} & 93.79 & 91.83 & 95.03 & -\\
     CodeBERT & 96.47 & 97.41 & 86.13 & 83.05 & 90.77 & 98.40\\
    
     \bottomrule
     \end{tabular}%
   \label{tbl:baselines}%
 \end{table*}%

An interesting finding on pre-trained models is that both pre-trained baselines perform very well on Java250 and Python800, but their accuracies are low on CodeNet C++ datasets, even though these two models are pre-trained on completely different corpora (CodeBERT is pre-trained on the CodeSearchNet dataset~\cite{husain2019codesearchnet} (without C/C++ code snippets), while C-BERT is pre-trained on Github C repositories). To further understand this unexpected phenomenon, we make an analysis of CodeNet datasets based on token sequences, which is demonstrated in Table~\ref{tbl:tokensequences}. In this table, we use code token sequences tokenized by the CodeBERT BPE tokenizer. We first list the lengths of token sequences, and it shows that the token sequences for C++ datasets are longer than Java/Python datasets. This brings a common weakness of most existing Transformer-based pre-trained models: these models require a fixed maximum input length due to the limitation of memory cost. For example, the maximum sequence length of CodeBERT and C-BERT is 512, which is shorter than the average sequence length of the C++1400 dataset. The overlength inputs are cropped to the maximum length and may harm the model from understanding the program's original semantics. We also make an analysis of the differences between each program classes at the token level. We treat the programs from the same class as a single text snippet by concatenating them and calculate the TF-IDF distances between every two classes. The average TF-IDF cosine similarity between all classes is reported in Table~\ref{tbl:tokensequences}. We can see that the average similarity scores of C++ datasets are higher, which means that in these datasets, programs from different classes are more similar at the token level compared to Java/Python datasets. As Transformer-based pre-trained models are conducted on token sequences, this "similarity" of tokens may hinder the models from learning the differences between program functionalities. In fact, as previous works \cite{sontakke2022code} have pointed out, Transformer models for code can be more sensitive to identifiers than logical structures, which further supports our findings.

 \begin{table}[htbp]
  \centering
  \caption{Statistics of CodeNet datasets in tokenized sequences.}
    \begin{tabular}{lcccc}
    \toprule
     & Java250 & Python800 & C++1000 & C++1400  \\
    \midrule
    Sequence length & 444.5 & 207.8 & 488.8 & 583.2 \\
    Average TF-IDF & 0.722 & 0.420 & 0.787 & 0.754  \\
    \bottomrule
    \end{tabular}%
  \label{tbl:tokensequences}%
  \vspace{-5mm}
\end{table}%


\subsection{RQ2: How does Tree-Transformer perform on node-level prediction tasks?}


Table~\ref{tbl:localization} demonstrates the results of wrong operator localization and repair. We report two accuracy metrics: localization accuracy and joint accuracy of localization and repair. We can find that compared with graph-structured baselines, Tree-Transformer achieves significantly better performance, especially on the joint loc\&rep accuracy. Our model gains an improvement of 7\% on joint accuracy compared with the best-performing baseline GIN. This indicates that our bi-directional propagation enables the model to learn effective node representations for node-level prediction tasks. 
The Transformer-based model for trees~\cite{shiv2019novel} performs poorly on this task, its accuracies lower than all GNN baselines. This suggests that changing the original tree structures by converting arbitrary trees to N-ary trees is harmful to node-level prediction. In the wrong operator dataset, the branching factor of 10\% syntax trees is larger than 15, so the structures of these trees are changed for this baseline.

In this task, the pre-trained model CodeBERT outperforms Tree-Transformer and achieves extremely high accuracy on both localization and repair. However, we must be aware that CodeBERT requires a large model size and pre-training on over 8 million data samples (while our model can be trained on a single GPU and does not require pre-training). A previous work\cite{kanade2020learning} has suggested that large pre-trained models can significantly outperform trained-from-scratch models and achieve very high results on synthesized localization \& repair tasks (e.g., variable misuse \cite{allamanis2018learning,vasic2019neural} and our WrongOp task). We believe this is because those pre-trained models excel in capturing the "naturalness" of input code from a large amount of data, so they easily recognize synthesized bugs that are "unnatural."


\begin{table}[htbp]
  \centering
  \caption{The accuracy of wrong operator localization and repair.}
    \begin{tabular}{lcc}
    \toprule
    Model & Loc Acc(\%) & Loc \& Rep Acc (\%)  \\
    \midrule
    GCN & 84.97 & 60.44 \\
    GIN & 85.69 & 61.61 \\
    GGNN & 84.77 & 58.71 \\
    Tree-PE~\cite{shiv2019novel} & 78.65 & 53.99 \\
    \midrule
    GREAT & 85.43 & 59.32 \\
    GCN+Graph & 86.37 & 65.05 \\
     GIN+Graph & 86.48 & 64.66 \\
     GGNN+Graph & 84.11 & 61.01 \\
    \midrule
    \textbf{Tree-Transformer (Ours)}   & 88.26 & 68.58 \\
    \midrule
    Transformer & 73.13 & 45.22 \\
    CodeBERT & \textbf{94.61} & \textbf{83.89} \\
    
    \bottomrule
    \end{tabular}%
  \label{tbl:localization}%
\end{table}%

The results for type inference are shown in Table~\ref{tbl:typeinference}. Tree-Transformer outperforms the GNN baselines with a large gap. This further demonstrates the strength of Tree-Transformer, as some predictions on variable types require long-dependency reasoning. For example, a user-defined type could be used long after its definition code block. Similar to the wrong operator dataset, Tree-Transformer outperforms vanilla Transformer on this type inference dataset, but the results are lower than the pre-trained model CodeBERT. The large performance gap between trained-from-scratch Transformer and pre-trained Transformer indicates that our Tree-Transformer may also benefit from pre-training techniques and further improve its performance. We will leave this as our future work.

\begin{table}[t!]
  \centering
  \caption{The accuracy of type inference on ManyTypes4TypeScript CST data.}
    \begin{tabular}{lc}
    \toprule
    Model & Accuracy(\%)  \\
    \midrule
    GCN & 46.35 \\
    GIN & 47.40 \\
    GGNN & 45.47 \\
    Tree-PE~\cite{shiv2019novel} & 51.71 \\
    \midrule
    GREAT & 52.45 \\
    GCN+Graph & 46.75  \\
     GIN+Graph & 47.91  \\
     GGNN+Graph & 45.71  \\
    \midrule
    \textbf{Tree-Transformer (Ours)}   & 54.78 \\
    \midrule
    Transformer & 49.54\\
    CodeBERT & \textbf{65.76}\\
    
    \bottomrule
    \end{tabular}%
  \label{tbl:typeinference}%
  \vspace{-5mm}
\end{table}%


\subsection{RQ3: How does each component of Tree-Transformer contribute to our performances?}
We perform an ablation study to further investigate the impact of each component in Tree-Transformer. Table~\ref{tbl-ablation} shows the results of Tree-Transformer when removing each component on program classification (Java250) and wrong operator localization. In this table, ``-" means removing a certain component from Tree-Transformer. The experiment of removing top-down propagation on wrong operator localization is omitted because leaf nodes cannot receive the information from their parent nodes only through bottom-up propagation. Note that after removing fraternal attention, the position encoding within it is also removed accordingly.
To separate position encodings from the fraternal attention, we add a new variant of Tree-Transformer where fraternal attention is removed, and position embedding vectors are added before the bottom-up parental attention.

From the results, first, we can see that our top-down propagation is succinct and powerful on both tasks. Without it, the performance drops greatly, which indicates that it is effective on both node-level prediction and tree-level classification tasks. Furthermore, even with only bottom-up propagation, Tree-Transformer still outperforms tree-structured baselines, showing that our attention-based neural network unit is effective. Moreover, modeling fraternal dependencies and sibling positions also contribute to both tasks. However, the effect of position encoding in Java250 is less significant than in WrongOp: removing position encodings in Java250 hardly affects the classification accuracy. This may indicate that in our program classification datasets, sibling order information is not essential for distinguishing programs of different classes. On the contrary, sibling orders in wrong operator localization are more important because locating wrong operators requires reasoning on relationships between operators and operands, and sibling dependency is a key part of these relationships.

An interesting finding is that only removing position encodings (keeping the fraternal attention) results in worse accuracies than jointly removing position encodings and fraternal attention on the WrongOp dataset. Because fraternal attention alone cannot learn from the sibling order information, adding this attention without given position information will deepen the model and make Tree-Transformer harder to train. If we keep the position encoding and only remove the fraternal attention, the results are still lower than the complete model. This suggests that position encoding works better when integrated with fraternal attention in Tree-Transformer.

\begin{table}[!t]
  \centering
  \caption{Ablation study on program classification (Java250) and wrong operator localization and repair. }
    \label{tbl-ablation}
    \begin{tabular}{lccc}
    \toprule
    \multirow{2}*{Model} & \multirow{2}*{Java250} & \multicolumn{2}{c}{WrongOp} \\
     & & Loc & Loc \& Rep \\
    \midrule
    Tree-Transformer & \textbf{95.32} & \textbf{88.26} & \textbf{68.58} \\
    -position encoding & 94.99 & 85.78 & 63.32\\
    -fraternal attention & 94.66 & 86.26 & 63.91\\
    -fraternal attention +position encoding & 94.95 & 87.18 & 65.64\\
    -top-down propagation & 94.01 & N/A & N/A\\  
    \bottomrule
    \end{tabular}%
  \label{tbl-discussion}%
\end{table}%

\section{Discussion}
In this section, we mainly discuss the strengths and limitations of Tree-Transformer compared with existing Transformer-for-code models, mainly from the perspective of efficiency. We also discuss the threats to validity of our work.

\subsection{Strengths}

The model structure of our Tree-Transformer is largely different from existing Transformer-based syntax tree modeling approaches: Tree-Transformer basically follows the tree-structured recursive formula \cite{goller1996learning,socher2011parsing}, while existing approaches \cite{zugner2021language,jiang2021treebert,peng2021integrating,tang2022ast} usually uses a sequential Transformer with node traversal sequences as inputs.

A main advantage of Tree-Transformer over sequential Transformer-based approaches is that Tree-Transformer can handle larger inputs. The core of Transformer is its multi-head attention mechanism. In a sequential Transformer encoder, the model performs self-attention on its input sequence. For a program syntax tree $\mathcal{T}$ with $N$ nodes, the time and memory cost of self-attention on sequential Transformers is $O({N}^2)$. This means that sequential Transformer on long sequences may take up a large amount of memory. To avoid out-of-memory problems, existing sequential approaches often set a not-so-long limit for their inputs, making them difficult to deal with long programs. For example, TPTrans \cite{peng2021integrating} assign a maximum number 512 for input CST leaf nodes. AST-Trans \cite{tang2022ast} requires the input ASTs not larger than 200 nodes. On the other hand, in Tree-Transformer, there are two multi-head attentions: the fraternal self-attention between siblings and a parent-children between a group of siblings and their mutual parent. Suppose the branching factor of $\mathcal{T}$ is $k$, then the memory cost of fraternal attention is $O({k}^2)$, and for parent-children attention is $O(k)$. In most syntax trees, $k\ll N$, so the memory cost of Tree-Transformer is significantly less than sequential Transformer. This allows Tree-Transformer to model syntax trees larger than the capacity of sequential Transformer-based approaches. For example, we do not apply any limitations on the syntax tree size for program classification and wrong operator localization/repair.

\subsection{Limitations}

The main limitation of Tree-Transformer, along with other recursive tree-structured models (e.g., Tree-LSTM), is that it runs slower than sequential Transformers or GNNs. In Tree-Transformer, before we update a node's state in the bottom-up pass, we must update all its children first and vice versa in the top-down pass. So for a syntax tree $\mathcal{T}$ with depth $\mathcal{D}$, the bottom-up/top-down Tree-Transformer unit must run for $2\mathcal{D}$ timesteps before we get the final node representations for $\mathcal{T}$. However, in sequential Transformer or GNNs with $L$ layers, their Transformer/message passing unit needs to run for $L$ times. For most program syntax trees and Transformer/GNN models, $L<2\mathcal{D}$ (e.g., our GREAT baseline only consists of 6 layers), so generally, the time efficiency of Tree-Transformer is lower than sequential Transformer/GNNs.

\subsection{Experiments on Efficiency}

 \begin{table*}[htbp]
  \centering
  \caption{Comparison of time and memory efficiency between Tree-Transformer, sequential Transformer and GIN.}
    \begin{tabular}{lcccccccc}
    \toprule
     & \multicolumn{2}{c}{GIN}  & \multicolumn{2}{c}{Transformer}  & \multicolumn{2}{c}{Tree-Transformer} &  \multicolumn{2}{c}{CodeBERT} \\
     Batch size & 256 & 1024 & 256 & 1024 & 256 & 1024 & 32 & 64 \\
    \midrule
    Training time (s/batch) & 0.13 & 0.45 & 0.42 & - & 1.09 & 2.53 & 0.61 & 1.17\\
    Training time (s/epoch) & 23 & 20 & 73 & - & 187 & 111 & 864 & 824\\
    Inference time (s/batch) & 0.05 & 0.17 & 0.12 & - & 0.51 & 0.86 & 0.21 & 0.41\\
    Inference time (s/epoch) & 2 & 2 & 7 & - & 30 & 13 & 99 & 97\\
    Memory (MB) & 15748 & 24348 & 36120 & OOM & 5729 & 18779 & 23821 & 46630\\
    \bottomrule
    \end{tabular}%
  \label{tbl:efficiency}%
\end{table*}%

To make a clear demonstration of the efficiency of Tree-Transformer, we compare the running time and memory occupation of Tree-Transformer with three baseline models: GIN, vanilla Transformer, and CodeBERT on the Java250 dataset. The results are shown in Table \ref{tbl:efficiency}. We can generally find out that the memory efficiency of Tree-Transformer is much higher than Transformer/GIN, while its time efficiency is lower. When the batch size is 256, which is the default setting in our experiments, the training time of Tree-Transformer is about four times as GIN and three times as Transformer, which is totally acceptable. As for the memory usage, Tree-Transformer only takes around $\frac{1}{5}$ of vanilla Transformer. When we increase the batch size to 1024, vanilla Transformer encounters the out-of-memory problem (we were using a GPU with 48GB memory), while Tree-Transformer only uses about 18GB memory. Generally, the memory effectiveness of Tree-Transformer makes it suitable for large input trees and large training batch sizes. As for the large pre-trained model CodeBERT, its training(fine-tuning) and inference time are both much longer than small-sized models, and its memory cost is higher even with a much smaller batch size.

\subsection{Threats to Validity}
The first threat involves the data processing step, which is the parsing of syntax trees in our experiments. To reduce this threat, we choose open source parsers which are evaluated by previous researchers on different downstream tasks. For the off-the-shelf syntax trees provided by the dataset creators, they are also parsed from widely-used parsers.

Another threat relates to the datasets we used in this paper. In the type inference task, we adopt the public ManyTypes4TypeScript benchmark, but filtered out some data samples to suit our experiments. This makes the dataset in this paper different from the public version. To reduce this threat, we ran all baseline approaches on the same filtered dataset as our Tree-Transformer. We will also make our preprocessing pipeline and preprocessed dataset public.

\section{Related Work}
In this section, we first briefly introduce existing tree-structured neural networks, then summarize their application in software engineering along with other structure-based deep learning approaches for code.

\subsection{Tree-Structured Neural Networks}
There has been long research interest in applying deep neural networks to tree-structured data. The earliest work on tree-structured neural networks is recursive neural network \cite{goller1996learning,socher2011parsing}. Recursive neural networks calculate the representation of a tree by accumulating node representations in a bottom-up manner. The recursive architecture further inspires later works \cite{tai2015improved,teng2017head,ahmed2019you,wang2020modular}. For example, Tai et al. \cite{tai2015improved} proposed Tree-LSTM, which uses an LSTM unit to replace the fully-connect neural network unit in recursive neural networks. 
Ahmed et al. \cite{ahmed2019you} proposed a recursive Transformer by performing self-attention on siblings. Although many classical tree-structured neural networks are proposed in the natural language processing community, in recent years, software engineering researchers have started to propose tree-structured models based on the programming language domain. For example, Mou et al. \cite{mou2016convolutional} proposed tree-based convolutional neural network (TBCNN). 
Bui et al. \cite{bui2021treecaps} further proposed TreeCaps, which extends TBCNN with capsule networks \cite{sabour2017dynamic} and achieved state-of-the-art results on program classification. 
Wang et al.~\cite{wang2020modular} further extended Tree-LSTM with different neural submodules for children state aggregation. 


\subsection{Modeling Code Structures with Neural Networks}
The above tree-structured neural networks have been widely applied in various software engineering tasks, such as code classification \cite{mou2016convolutional,zhang2019novel,bui2021treecaps}, code summarization \cite{wan2018improving,bui2021treecaps}, code search \cite{wan2019multi}, etc. In recent years, as the popularity of graph neural networks grows, researchers started to apply GNN on programs. Some works directly use GNNs to model program syntax trees \cite{yao2021learning,puri2021project}, while other works tried to augment syntax trees with additional hand-crafted edges \cite{allamanis2018learning,wang2020detecting} or leverage control-flow/data-flow graphs \cite{zhou2019devign,wan2019multi,xue2018appdna,li2020appdna,ma2022graphcode2vec}. 

Some other works tried to extract substructures from syntax trees, and use sequential (or partly sequential) neural networks to model code substructures. For example, Zhang et al. \cite{zhang2019novel} proposed ASTNN, which decomposes an AST into a sequence of subtrees following the order of statements, and combines GRU and recursive neural network to encode the subtree sequence. Alon et al. \cite{alon2019code2vec,alon2018code2seq} sample paths between nodes from ASTs, and use feed-forward neural networks/LSTM to encode the set of AST paths.

Recently there have been some attempts to integrate the tree/graph structures of code into the popular Transformer model. Most of these works manipulate the attention mechanism or position encoding of Transformers to provide the model with structural information\cite{shiv2019novel,li2021improving,zugner2021language,peng2021integrating}. For example, Shiv et al. \cite{shiv2019novel} proposed a novel position encoding technique to extend Transformers to N-ary trees, and evaluate their approach on a code translation task. Hellendoorn et al. \cite{hellendoorn2019global} proposed Graph Relational Embedding Attention
Transformers (GREAT) to model program graphs. Guo et al. \cite{guo2020graphcodebert} proposed a pre-trained model GraphCodeBERT, which integrates simplified data flow structure into code token sequences.
Z\"{u}gner et al. \cite{zugner2021language} proposed Code Transformer, which integrates multiple distance metrics of AST nodes into the relative position encoding of Transformers. 
Jiang et al.~\cite{jiang2021treebert} proposed TreeBERT, a Transformer-based pre-trained model on ASTs. TreeBERT extracts paths from ASTs as the inputs to the transformer model, and proposed a novel tree-based position encoding for nodes.
Peng et al. \cite{peng2021integrating} proposed TPTrans, which integrates the information of paths in CSTs by encoding the paths as the position encodings in a Transformer model.
Guo et al. \cite{guo2022unixcoder} proposed a pre-trained model Unixcoder, which uses the structure information of ASTs in the pre-training stage. However, the tree structure is not used in the fine-tuning/inference stage.

\section{Conclution}
In this paper, we propose a novel tree-structured network: Tree-Transformer for program representation learning. Tree-Transformer leverages the powerful multi-head attention in two dimensions: fraternal and parental, to capture the dependencies between siblings and ancestors/predecessors on trees. Motivated by the neglect of the global dependency among nodes in current recursive-structured neural networks or GNNs, we propose bidirectional information propagation along trees and extend existing recursive neural network architecture with a novel top-down unit. Experiments on three different tasks: program classification, wrong operator localization\&repair, and type inference have demonstrated the effectiveness of our proposed approach. In the future, we would like to further explore the potential of our Tree-Transformer with pre-training techniques. Another possible improvement is to integrate Tree-Transformer with GNNs so that the model can simultaneously handle syntactic structure information and control/data flows.

\section*{Acknowledgments}
This research is partially supported by the National Research Foundation, Prime Ministers Office, Singapore, under its National Cybersecurity R\&D Program (Award No. NRF2018NCR-NCR005-0001), NRF Investigatorship NRF-NRFI06-2020-0001,  the National Research Foundation through its National Satellite of Excellence in Trustworthy Software Systems (NSOE-TSS) project under the National Cybersecurity R\&D (NCR) Grant award no. NRF2018NCR-NSOE003-0001, the National Research Foundation Singapore and DSO National Laboratories under the AI Singapore Programme (AISG Award No: AISG2-RP-2020-019).

\bibliographystyle{IEEEtran}
\bibliography{IEEEexample}


\end{document}